# AI-assisted prediction of catalytically reactive hotspots in nanoalloys

Jolla Kullgren[1], Peter Broqvist[1], Ageo Meier de Andrade[1], Yunqi Shao[1], Seungchul Kim[2], Kwang-Ryeol Lee[2]

[1] Department of Chemistry, Ångström Laboratory, Uppsala University, 75121 Uppsala, Sweden

[2] Computational Science Research Center, Korea Institute of Science and Technology (KIST), Hwarangno 14-gil 5, Seongbuk-gu, Seoul 136-791, Republic of Korea

We have developed a computational framework to create reactivity maps for nanoalloy systems. The framework is accelerated by ML techniques that are used for both rapid homotop optimization, as well as rapid construction of the corresponding reactivity maps.

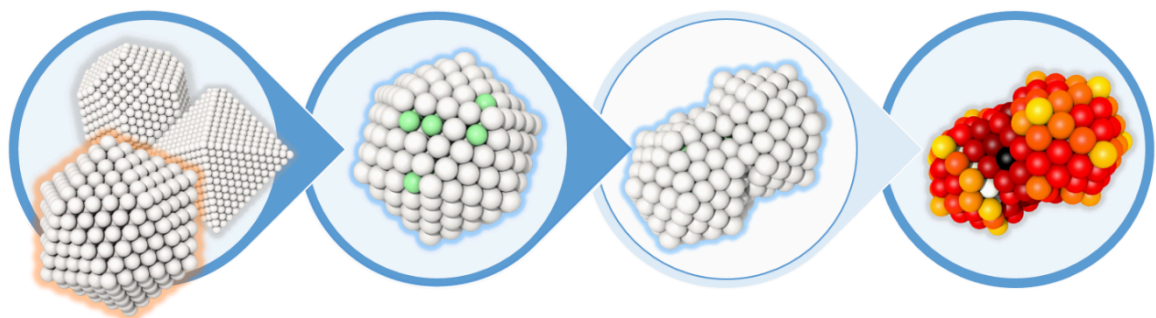

**Fig 1**. An overview of the computational approach developed in the current article.

*Nanoalloys offer the possibility of tailoring chemical properties by changing their composition, as well as the shape and size. This flexibility comes with scientific challenges that cannot be tackled using experiments alone. Moreover, standard theoretical methodologies are often found to be too slow to map the broad structural diversity that nanoalloys could span [1]. One example is to identify reactive hotspots in bimetallic catalysts, which are needed to tailor future heterogeneous catalysts. In the present work, we have developed an AI-assisted prediction of catalytically reactive hotspots in nanoalloys. The first step towards understanding the catalytic reactivity of metallic nanoalloys is to find thermodynamically stable atomic arrangements, i.e., finding the most stable nanoparticle homotop. To accelerate this step, a fast screening method is required. In this work, we use the Metropolis Monte Carlo method combined with a lattice Machine Learning Potential [2] (ML) trained on the second Nearest Neighbors Modified Embedded-Atom Method Potential [3] (2NN-MEAM) data, here demonstrated for various $Pt_3Ni$ homotops. The result from such a screening procedure is a nanoparticle showing a core-shell-like geometry, with a Ni core and a Pt surface, which is decorated with few Ni atoms (see figure (b) below). In the proposed screening method, the ML potential replaces the need for ionic relaxation between sequential Monte Carlo steps, which accelerates the prediction of the optimized geometries significantly. The second step concerns the prediction of reactive hotspots. In this step, we use the fact that for metallic nanoalloys, there is a clear correlation between catalytic activity and the so-called metal d-band centers [4]. However, electronic structure calculations based on, e.g., the density functional theory (DFT), for large nanoalloys are computationally costly. In our scheme, we instead use a sequential multiscale modeling approach where we, based on DFT data,*

*generate density functional tight-binding (SCC-DFTB) parameters that enable a fast computation of accurate d-band centers in bulk and nanoparticles of Pd and Ni. To allow for even larger nanoalloys, we use the SCC-DFTB method to create extensive training sets to train an ML potential again. The result from this step is very efficient, i.e., fast, accurate and scalable (both when it comes to size and composition) AI model that allows for the mapping of d-band centers in nanoalloy. The framework has been proved to be successful in the modeling $Pt_3Ni$ nanoalloys of experimental size, see figure below, and can be readily extended to any other nanoalloy system.*

# 1 INTRODUCTION

Peculiar and dramatic chemical effects have been found to emerge merely as a consequence of the reduction of the particle size of nanoparticles (NPs). This has instilled both intense research into their fundamental properties and intense efforts towards new technological applications, not least within heterogeneous catalysis, the importance of which cannot emphasised enough. The properties of metal NPs, can be tailored by controlling their size, shape, and by alloying. Alloy systems can be classified into intermetallic compounds and solid-solutions depending on whether they display long-range order or not. One key reaction when it comes to metal NPs is the oxygen reduction reaction (ORR) in which molecular oxygen is reduced to either water or hydrogen peroxide ($H_2O_2$). Here, the platinum-group metal (PGM) NPs hold a special place due to their high activity and stability. Consequently, PGM NPs are crucial components in technologies reliant on an efficient ORR, some of which are likely to play a central part in the transition to a future fossil-free economy. For example, Pt-based nanoparticles are key components in **fuel cells** where they promote the reduction of molecular oxygen at the cathode, while Pd-based NPs can be used to produce $H_2O_2$, one of the top 100 most important chemical compounds, directly from gaseous $H_2$ and $O_2$. The high cost of PGM remain a major bottleneck in large scale commercialization of technologies based on them[3]. This fact have inspired much research into avenues to reduce the amount or even completely replace PGM in ORR catalysts, so-called low- or PGM-free catalysts, which is a hot topic today. Many of these strategies include NP tailoringAs mentioned, this can be combined with alloying.

A particularly interesting class of NPs are the Pt-based alloy-core-shell particles with Pt-covering the surface, forming a so-called Pt-skin. A popular example is the intermetallic $Pt_3Ni$ catalyst with a 90-fold increase in ORR activity compared to commercial platinum/carbon catalysts[4]. In fact, the activity of the catalyst already comply with the U.S. Department of Energy **(DoE)** 2020 target but the durability is still insufficient[5]. Adding a third, a fourth or even a fifth metal to the mix could improve on the stability. Promising results have been seen for e.g. $Pt_2CuNi$[6]. At least two recent reviews points to the pivotal role of simulations in the search for new low-PGM catalysts stating that:

> *"[…] computational materials design seems likely to become more important as the materials become more complex with increasing numbers of elements and different structures."* [7]

> *"However, voluminous studies on the durability of crucial materials are still insufficient. In particular, studies on improving the durability using quantum mechanics […]"*[5]

Developing the necessary simulation tools to take on such a task is challenging! The complexity and scale of the systems require multi-scale modelling involving parametrized methods. The inherent many-body character of metallic bonding leads to a large increase in the number of parameters as more elements are added. A proper account of the electronic structure in these complex system will also be required to make reliable predictions of activity and stability.

Numerous descriptors to predict and describe reactivity (adsorption energies, reaction barrier etc.) of individual sites on the surface of metal particles have been presented in the literature. Recent examples include those based on generalized coordination number[5] or orbitalwise coordination number[6]. Nevertheless, the seminal d-band model of Hammer and Nørskov[4] remain one of the most widely applied descriptors owing to the fact that d-band centers can be readily calculated with DFT without any additional parameters. The same fact also means that it is difficult to apply the model to large, and complex systems, like nanoparticles. In the current contribution, we make use of an approximate DFT approach, self-consistent charge density functional tight binding (SCC-DFTB) to target systems beyond the reach of routine DFT simulations. The DFTB data is used to train a machine learning (ML) model that allows for fast prediction of d-band centers in even larger nanoalloy systems.

# 2 METHODS

### 2.1.1 Electronic structure calculations
All plane wave basis DFT simulations are performed using Vienna Ab initio Software Package (VASP[7–10]) using the Projector Augmented Wave (PAW[11,12]) method for pseudo potentials. The plane-wave basis-set was truncated at 600 eV.

All density functional tight binding calculations were performed using the DFTB+ software[13,14].

### 2.1.2 Structural models
We consider particles in three general shapes: cuboctaheral, octahedral and icosahedral. All particles have a composition being as close as possible to a 3:1 ratio of Pt vs Ni, i.e. close to a $Pt_3M$ stoichiometry.

### 2.1.3 Homotop optimization using Monte Carlo
Our protocol for generating nanoalloys is the following. A particle of a given shape and size is generated from the perfect fcc bulk structure and atoms are randomly distributed over the nanoparticle "lattice". To obtain a representative homotop, the particle is subject to a Metropolis Monte-Carlo (MC) simulation, using the canonical ensemble and a temperature of 300 K, were atoms are redistributed over the lattice by sequential swapping of atoms. Each trail move involves a full relaxation of the structure, until the forces on each atom is less than 0.05 eV/Å, before evaluating the energy difference between initial and final states in any attempted swap. We found that the relaxation step was crucial, since omitting it from the MC simulation led to homotops of much higher energy, even after a local geometry optimization of the final structure. Energies are evaluated using the 2NN-MEAM method by Lee and Baskes[15] using parameters of Kim et al.[16] Each Monte-Carlo simulation consisted of 10'000 attempted swaps and were performed using an in-house python code.

To allow for efficient homotop optimization in large nanoparticle models we have trained an on-lattice ML potential using a training-set derived with the 2NN-MEAM potential. We use the PiNN code. The training to validation ration was 80/20, we used a cutoff of 3.5Å and a gaussian basis. We used a learning rate of $3 \cdot 10^{-4}$, batch size of 10. All optimization was performed using 3 '000' 000 steps.

#### 2.1.4 Interface modelling

Models for interfaces between two nanoparticles were constructed using an in-house *python* code for automated assembly of interfaces. The code is based on the quick hull algorithm by Barber *et al.*[17] which allows for, among other things, rapid identification of convex hulls using a set of points as input, in the current case the coordinates of the atoms in a nanoparticle model. The procedure to construct interfaces is as follows:

- Facets are constructed by merging simplicies of the convex hull whose surface normal forms an arcus cosine of 0.95 or larger. This is done for each particle individually.
- The two particles are aligned such that the normal of their largest facet are pointing towards each other and are parallel to the x-axis.
- The geometrical mean of the two facets are placed at a distance corresponding to the nearest neighbour distance, $r_{NN}$ in the optimized bulk Pt phase from each other along the x-axis. The geometry is fully relaxed using the 2NN-MEAM method
- The procedure is repeated with the particles being rotated with respect to one another around the x-axis using angles between 0 and 180 in steps of 15. For each rotation 12 additional interface geometries, with the particles also being displaced in the yz-plane on a circle with radius $r_{NN}$, using angles between 0 and 360 in steps of 30, are generated.

This procedure leads to a total of 156 interface geometries for each particle pair. Each geometry is optimized using the 2NN-MEAM method and the most stable configuration is selected as a representative interface structure.

# 3 RESULTS

The aim of the current investigation is to develop a computational approach to construct reactivity maps (d-band center maps) for nanoalloys and nanoalloy interfaces of "experimental size". The layout of our computational approach is given in **Fig. 1**. We will primarily show data for particles of icosahedral shape but selected results from cuboctahedral and octahedral will also be shown.

## 3.1 STRUCTURE

### 3.1.1 Homotop optimization

**Fig. 2** show the optimized homotops using the 2NN-MEAM-based MC for the cuboctaheral, octahedral and icosahedral $Pt_3M$ particles with 146-147 atoms. We observe the formation of an almost complete Pt-skin in all of the shapes although corners and edges of the cuboctahedra and octahedral tend to be decorated with Ni atoms (see **Fig. 2**). Similar behaviour is observed in particles up to 923 atoms, which were the largest that were optimized with the 2NN-MEAM-based MC. The Pt-skin formation is in line with findings in previous work in the literature[18].

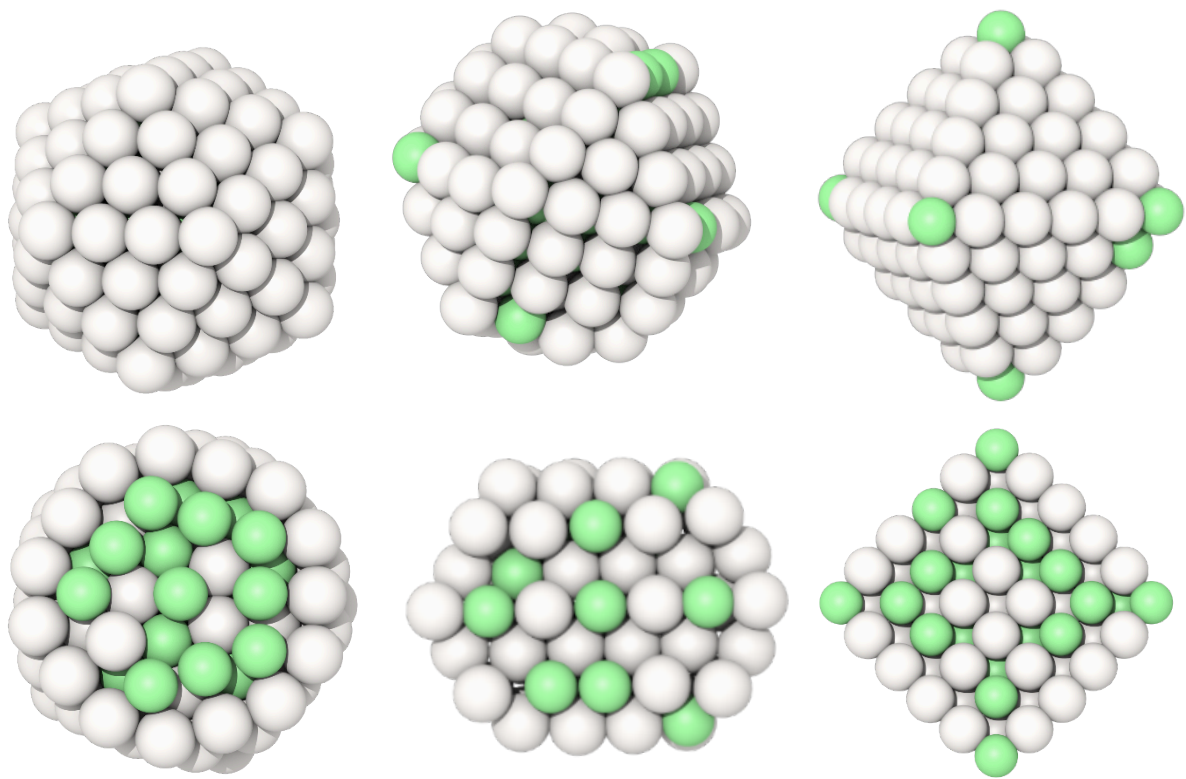

    **a) Icosahedra**        **b) Cuboctahedra**        **c) Octahedra**

*Fig. 2. Optimized homotops emerging after 2NN-MEAM based MC simulations for an icosahedral Pt$_3$Ni particle with 147 atoms (a), a cuboctahedral Pt$_3$Ni particle with 147 atoms (b), and octahedral Pt$_3$Ni particle with 146 atoms (c). Top panel show the full structures and the bottom panel show slices going through the middle of the particles. Pt and Ni atoms are represented by gray and green spheres, respectively.*

### 3.1.2 Accelerated homotop optimization using ML

In order to accelerate homotop optimization in larger particles, we have developed an on-lattice ML potential for the set of icosahedral particles that allows us to calculate homotop energies without having to perform time consuming geometry optimizations.

We use an iterative process (an active learning protocol) to fit the ML potential. We obtain a first potential (generation I) from fitting to a set of 400 random homotops, 100 structures each from particles of sizes 147, 309, 561 and 923. The ML method is presented with the unrelaxed geometries together with the energies obtained after a full geometry relaxation using the 2NN-MEAM method. The quality of the generation I ML potential can be appreciated by looking at **Fig. 3**, which give the correlation plot between the target energies and those predicted by the model. Clearly, the potential is capable of describing this relative energy in the set of random homotops. In a second step, we perform a homotop optimization using the ML potential for the four sizes of icosahedra present in the training-set. Snapshots recorded every hundred steps in these MC runs are fully optimised at the 2NN-MEAN level and appended to the training-set (400 new structures). The ML potential is refined using the augmented training-set (generation II). New MC runs are performed and used to augment the training-set further yielding ML potentials of generation III and IV.

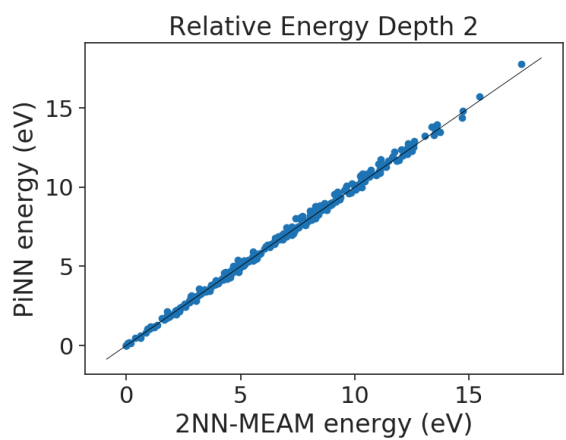

*Fig. 3*. Correlation plot of predicted energies of random homotops using ML on-lattice models versus the values after full relaxation at the 2NN-MEAM level of theory.

### 3.1.3 Interface modelling

We have constructed a number of nanoalloy interfaces using pairs of equally size icosahedral particles and the technique described in section **2.1.4**. Particles up to 1415 atoms were used in these simulation, and in all cases the most stable interfaces are coherent, i.e. forming a continuation of the bulk lattice across the interface, see **Fig. 4**.

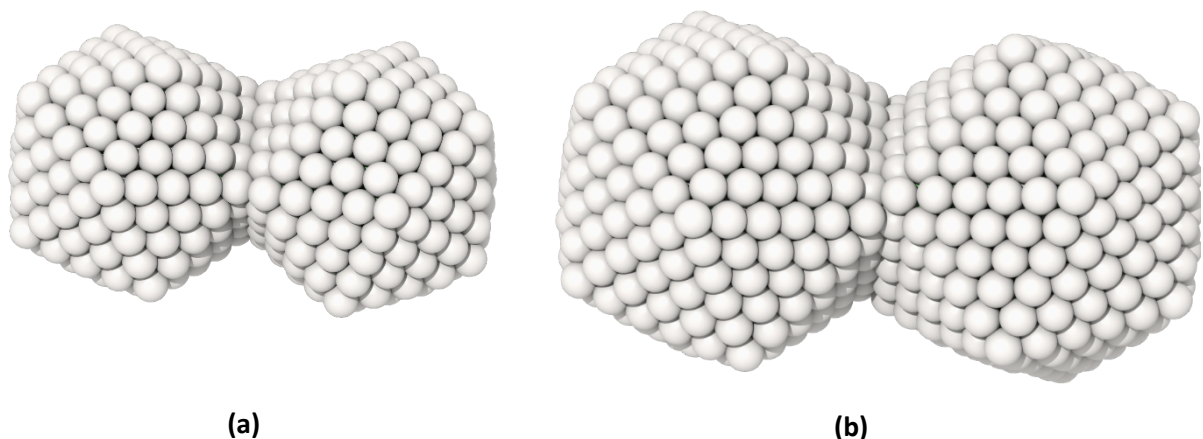

(a)                                                                 (b)

*Fig. 4*. Optimized interface structure for an assembly of two 309 atom icosahedra (a), and two 561 atoms icosahedra (b), see text for more details.

## 3.2 REACTIVITY MAPS

Before turning to our construction of reactivity (d-band) maps, we first describe our DFTB parametrization and evaluate its quality in terms of describing d-band centers compared to DFT.

### 3.2.1 DFTB parametrization and validation

The parameters were fitted by minimizing the deviation between the d-band center in bulk Pt, bulk Ni and a simple bulk PtNi structure calculated at the DFT and DFTB level of theory, respectively. **Fig. 5** compares calculated density of states and d-band centers from DFT to DFTB method with low and high compression of the wavefunctions and electron density, respectively. Using a low compression, the d-band centers calculated with DFTB are within 0.1 eV from those at the DFT level.

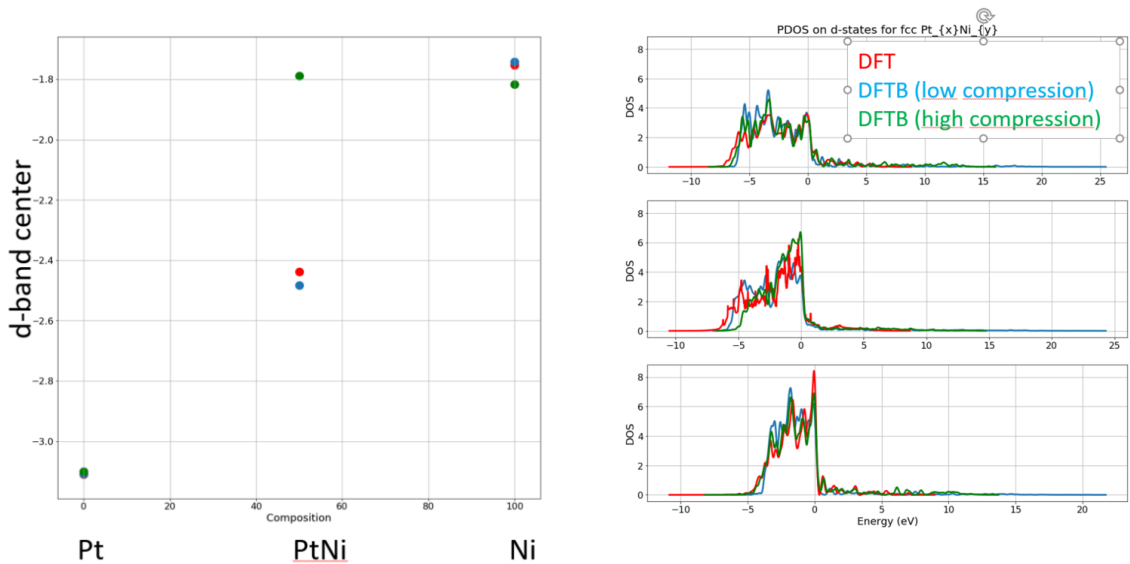

*Fig. 5*. Calculated d-band centers (left panel) and density of states (DOS, right panel) from DFTB vs DFT for bulk fcc Pt-Ni alloys. Two different compressions for DFTB are shown.

As a further validation of our DFTB parameterization, we compare calculated d-band centers in a 147 atom icosahedral particle. We used a randomly generated homotop with a geometry optimized at the 2NN-MEAM level of theory. We have calculated the local d-band centers for all atoms at the DFT and DFTB level of theory. **Fig. 6**a shows reactivity maps, in the form of a colormap projected on the atoms, calculated using DFT. The maximum deviation between the d-band centers calculated using DFTB and DFT is 0.15 eV for the surface atoms and 0.4 eV for the bulk atoms (see **Fig. 6**b).

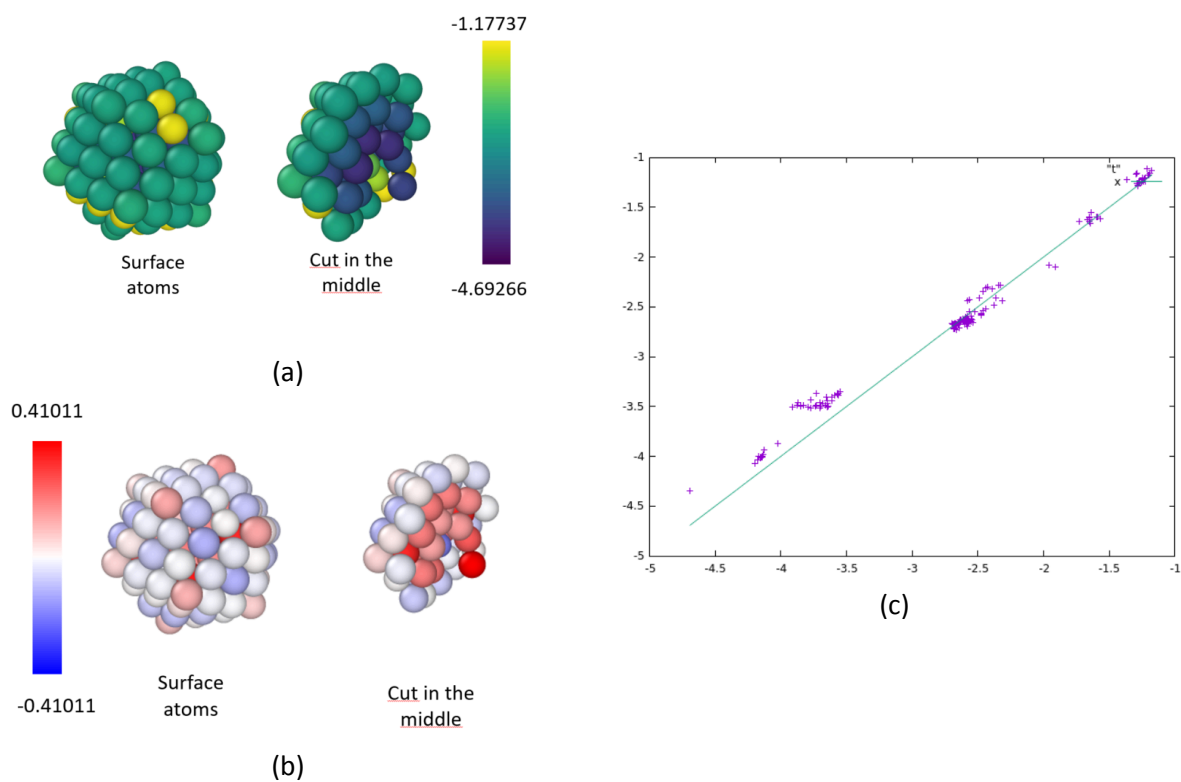

*Fig. 6. (a) d-band center calculated at the DFT level of theory on a nano particle geometry optimized at the 2NN-MEAN level of theory. (b) The difference in d-band center when calculated at the SCC-DFTB level instead of DFT level using the same particle. (c) correlation plot with DFT d-centers on x-axis and DFTB d-centers on y-axis.*

### 3.2.2 Reactivity maps using ML

In order to accelerate the generation of reactivity maps and to allow us to target larger particle sizes, we trained three ML models to predict DFTB d-band centers. The three ML methods have depth 1, 2 and 3, respectively. Depth 1 corresponds to taking information form only the nearest neighbours and is therefore, in a sense, akin to using coordination numbers. Depths 2 could in the same fashion be related to the generalised coordination number. Depth 3 goes beyond… The data-set used in the fitting consists of 194'000 unique d-band centers from the 400 icosahedral particles of up to 1415 atoms in size from the generation I data-set used to fit our lattice ML model. The ability of the ML models to predict DFTB d-band centers is illustrated in **Fig. 7**. There is a clear improvement in the prediction when increasing the depth from 1 to 2 to 3. We also note that errors in the estimation of the DFTB d-band centres using our ML methods are much smaller than the difference between DFT and DFTB d-centers, thus the ML replacement does not deteriorate the DFTB prediction any further. We therefore expect that surface d-band centers are predicted within about 0.15 eV compared to the DFT reference using our ML approach.

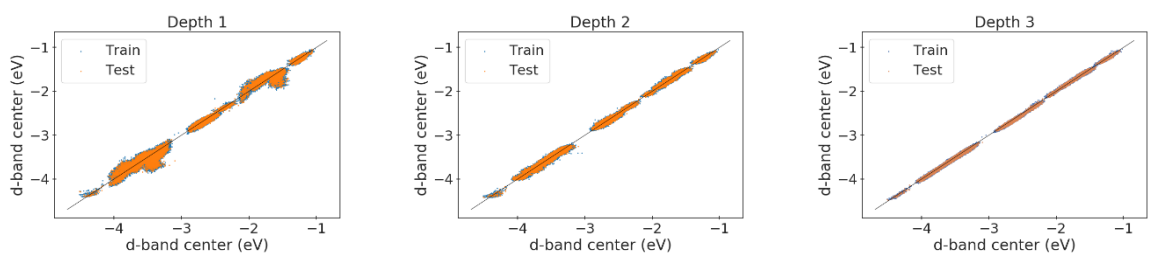

*Fig. 7. Correlation plot of d-band center predicted with ML versus the values calculated at the SCC-DFTB level of theory.*

### 3.3 REACTIVITY MAPS OF REALISTIC PARTICLES

Finally, to take real advantage of our machinery, we have simulated reactivity maps for icosahedral Pt$_3$Ni nanoparticles up to 1415 atoms in size and also considered the impact of an interface between them. The only required input to perform such simulations using our platform is the desired size of the particle, expressed in the number of atoms. An example is shown in **Fig. 8**. The apex atoms clearly sticks out having a d-center at higher energy than the other atoms at the surface. We also note that the d-band centers close to the interface are more negative. Thus, depending on the type of reaction, these different types of atoms could become reactive hot-spots.

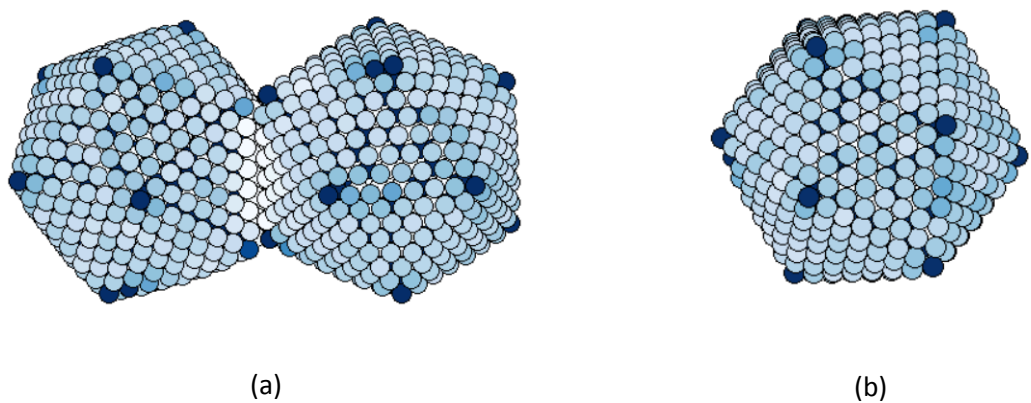

(a)          (b)

*Fig 8. (a) A nanoparticle interface between two 1415 atoms icosahedral nanoparticles. (b) isolated icosahedra nanoparticle. The color show the ML predicted d-band centers at the surface. The dark atom close to the apex positions of the right hand particle in the interface structure is an Ni atom breaking through the Pt-skin.*

# 4 Conclusions

We have developed a scalable and efficient AI-assisted framework for predicting catalytically active sites in nanoalloys. By combining a lattice-based machine learning potential with Metropolis Monte Carlo sampling, we rapidly identified thermodynamically favorable nanoparticle homotops, as demonstrated for $Pt_3Ni$ systems. The predicted core–shell structures serve as input for a multiscale modeling approach that leverages SCC-DFTB and machine learning to accurately map d-band centers, key indicators of catalytic activity. This method circumvents the computational cost of traditional DFT calculations and enables high-throughput screening of large and compositionally diverse nanoalloys. Our approach has been validated on experimentally relevant $Pt_3Ni$ particles and can be generalized to other alloy systems for the rational design of next-generation catalysts.

# 5 Acknowledgments

We would like to acknowledge the National Academic Infrastructure for Supercomputing in Sweden (NAISS) for computing resources and STINT for supporting the collaboration between Sweden and Korea.